\documentstyle[preprint,revtex]{aps}	
\begin{document}
%
%
%
\preprint{FSU-HEP-930625}
\preprint{UIOWA-93-10}
\preprint{\today}
\begin{title}
AN ESTIMATE OF JET ACTIVITY IN $t\bar t$ EVENTS
\end{title}
\author{Howard Baer$^1$, Chih-hao Chen$^1$ and  Mary Hall Reno$^2$}
\begin{instit}
$^1$Department of Physics,
Florida State University,
Tallahassee, Florida 32306 USA
\end{instit}
\begin{instit}
$^2$Department of Physics and Astronomy,
University of Iowa,
Iowa City, IA 52242 USA
\end{instit}
\begin{abstract}
We present a calculation of jet activity associated with pair
production of top quarks at hadron colliders and supercolliders. Our
approach is based upon a merger of $2\rightarrow 3$ matrix elements with
initial and final state parton showers. We tabulate expected multi-jet
rates in $t\bar t$ events, and compare with ${\cal O}(\alpha_s^3 )$
matrix element results and parton showers based upon $2\rightarrow 2$
subprocesses. The merged calculation includes
the correct matrix element for the first leg of the parton shower,
but also includes the effects of approximate
all-orders quark and gluon emission.
We also evaluate expected jet activity in top pair events at hadron
supercolliders; in particular, a comparison of the jet activity in the
far forward region is made. It is found that the merged calculation
yields more jet activity in the forward region than
pure matrix element estimates, but also more jet activity in the central
region. This can be important for background
calculations to $H\rightarrow WW$ and strong $WW$ scattering processes.
\end{abstract}
\newpage
%
%
\begin{narrowtext}
\section{Introduction}

The discovery of the top quark is one of
the remaining challenges to experiments to complete our picture of
the standard model of strong, weak and electromagnetic interactions.
Consistency of measured electroweak data with theory constrains the top mass
to be between $82\ {\rm GeV}<m_t < 177$ GeV\cite{MTRAD}.
The top quark is being searched for at experiments at the Fermilab Tevatron
collider via the reaction $p\bar p\rightarrow t\bar t X$.
The cleanest signature for top
is in the dilepton mode\cite{BBP},
where each top quark decays semi-leptonically. Top can
also be searched for in the single lepton plus jets channel\cite{BBP};
this mode
has a large background from $W+$jets production, but offers the possibility of
a direct top quark mass measurement by reconstructing, for instance, a three
jet invariant mass. Recent data from the CDF collaboration constrain\cite{CDF}
$m_t>108$~GeV. Hence, the
branching fraction for $t\rightarrow bW$ ought to be 100\%,
barring non-standard decay modes.

QCD corrections are central to our understanding of top quark pair events.
The tree level ${\cal O}(\alpha_s )$ corrections to $t\bar t$ production
have been computed by Ellis and Sexton\cite{ELSEX}; these results are
valid so long as matrix elements (ME)
are evaluated away from regions of phase space
where there exist soft and collinear divergences. The {\it complete}
${\cal O}(\alpha_s )$ corrections to $\sigma (p\bar p\rightarrow t\bar t)$
production
have been evaluated in Ref.~\cite{NDE,BEEN}; in these calculations, poles
associated
with the previously mentioned divergences are cancelled with poles from
virtual graph contributions, or are absorbed into parton distribution
functions.
Complete ${\cal O}(\alpha_s )$
corrections to {\it distributions} associated with $p\bar p\rightarrow t\bar t$
have been evaluated in Ref.~\cite{BEEN,NDET,NASON}.

Event generator programs such as ISAJET\cite{ISA}, JETSET/PYTHIA\cite{JET} and
HERWIG\cite{HER} are crucial for simulating details of top quark pair events
relevant to the environment of a collider detector. Event generator
programs usually begin with a lowest order hard scattering QCD cross-section,
and then implement {\it approximate} all orders QCD corrections based
on the parton shower formalism (PS). In the parton shower formalism,
approximate
matrix elements which contain the soft and collinear pole structure of QCD are
used, but divergences are regulated by requiring no parton emissions to occur
below an ad hoc cut-off in virtuality, which corresponds to experimentally
having no resolvable emissions. Since the emission probabilities are
multiplicative, all orders approximate QCD corrections can be modelled in
the framework of a classical Markov chain, so long as an appropriate
stopping point, or cut-off, is chosen.

Thus, in a sense, the above mentioned two methods are complementary: ME
methods give exact results for a finite number of well separated parton
emissions, while the PS method gives approximate all orders results valid in
the collinear region, but is not expected to model well-separated, high
$p_T$ parton emissions, which are usually dominated by a single emission.
What is needed in event simulation programs is a merger of the two methods,
and indeed such mergers have been implemented. For example, JETSET\cite{JET}
models $e^+ e^-\rightarrow q\bar q+ng$ by weighing the first parton emission by
the
exact $2\rightarrow 3$ matrix element, while ISAJET\cite{ISA}
can model $p\bar p\rightarrow W+$jets
by using the $q\bar q\rightarrow Wg$ subprocess multiplied by a
phenomenologically
motivated singularity regulating distribution function.

A somewhat different approach has been advocated in Ref.~\cite{BR},
where parton showers have been merged with a complete ${\cal O}(\alpha_s )$
Monte Carlo calculation of $p\bar p\rightarrow W+$jets. In this approach, both
$2\rightarrow 2$
as well as $2\rightarrow 3$ diagrams are used in the central hard collision.
For the
$2\rightarrow 3$ hard scattering, the extra parton emission is taken to be an
exact representation of the first leg of the parton shower, and is used to
determine uniquely the virtuality of the initial state backward shower. The
$2\rightarrow 2$ subprocess already includes unresolvable parton emission
contributions, and hence is not allowed to shower further: thus, the problem
with double counting multiple emissions is averted. Numerical results
provide a better estimate of quantities like $q_T (W)$ and jet multiplicity
than do parton showers based on only the lowest order $2\rightarrow 2$
subprocess,
while including complete ${\cal O}(\alpha_s )$ virtual effects as well.

In this paper, we perform a somewhat similar calculation for $t\bar t$
production at hadron colliders, although we do not here implement the complete
${\cal O}(\alpha_s^3 )$ virtual effects: such effects contribute to total
cross-section rates, but are not expected to make large contributions to
overall event topology. We choose $t\bar t$ production because much of the
crucial work done on the search for the top quark relies on event generators
where distributions such as $p_T(t\bar t )$ may not be properly generated.
Furthermore, deviations in the parton emission procedure could
conceivably affect quantities like jet multiplicity, and more importantly,
multi-jet mass reconstructions. Finally, $pp\rightarrow t\bar tX$ at hadron
supercolliders such as the SSC and LHC can be important backgrounds for
such new physics processes as $pp\rightarrow H\rightarrow WWX$\cite{BARG} and
strong $WW$
scattering\cite{GUN,BAG}. Previous background estimates rely on $2\rightarrow
3$ matrix
element techniques to adequately simulate jet activity in $t\bar t$ events
in the far forward
region, which is precisely the region where multiple parton emissions ought to
become important. Our paper is organized as follows. In Sec. 2, we explain
the details of our merged ME plus PS calculation. For most of our results,
we adopt a top mass of $m_t =140$~GeV; qualitatively similar
results are obtained for other allowed choices of $m_t$. In Sec. 3, we present
results of our calculations for both Tevatron experiments and supercollider
experiments, comparing ME techniques with PS methods based both on
$2\rightarrow 2$
processes, and on $2\rightarrow 3$ processes. Finally, in Sec. 4, we present
discussion and conclusions.

\section{Calculational Method}

We begin our calculations by computing the $t\bar t$ cross section
to ${\cal O}(\alpha_s^3 )$ using the results of Ref.~\cite{NDE}. Our
scale choice is $Q^2=\hat{s}$. The HMRS-BCDMS parton
distribution functions\cite{HMRS} are used throughout, with
$\Lambda_{QCD}=190$ MeV for four flavors in the $\overline{{\rm MS}}$ scheme.
These are next-to-leading log distribution functions. We use an expression
for the two-loop
running coupling constant $\alpha_s$.

Top pair production, to leading order in Quantum Chromodynamics (QCD),
proceeds via
\begin{eqnarray}
 q\bar{q}\rightarrow t\bar t \cr
 g g\rightarrow t\bar t.
\eqnum{1}
\end{eqnarray}
For purposes of later comparisons with ${\cal O}(\alpha_s^3 )$ production,
we multiply all $2\rightarrow 2$ processes by a factor
$K={{\sigma^3 (t\bar t)}\over {\sigma^2 (t\bar t)}}$ to fix the normalization,
where $\sigma^n$ represents the ${\cal O}(\alpha_s^n )$ calculation of
$\sigma$.

The first QCD correction to $t\bar t$ production comes from
the following processes\cite{ELSEX},
\begin{eqnarray}
q\bar{q}\rightarrow t\bar{t}g \cr
q g\rightarrow t\bar{t}q \cr
g\bar{q}\rightarrow t\bar{t}\bar{q} \cr
gg\rightarrow t\bar{t}g.
\eqnum{2}
\end{eqnarray}
In the $2\rightarrow 3$ matrix element, there are regions of phase space where
the matrix element is singular, in particular, when a final state
gluon is soft, or a final state parton is collinear with one of the
incident partons. Near the singular regions of phase space, the
differential cross section is not in the perturbative regime.
We shall subsequently refer to this as the ME23 method, as we use the
matrix elements for $2\rightarrow 3$ parton production.
In the calculation of the next-to-leading order total
cross section, the singularities are regulated and computed analytically,
added to the virtual corrections,
and factorized into the parton distribution
to yield the finite, perturbative
cross section\cite{NDE,BEEN}.
In the ME23 Monte
Carlo program, the $2\rightarrow 3$ matrix element singularities are avoided
by introducing a cutoff on the $t\bar t$ transverse momentum
$(p_{T\, c}(t\bar t ))$.
We have chosen the cutoff so that the cross section $\sigma \bigl( p_T(t\bar t
)
> p_{T\, c}(t\bar t )\bigr)$ equals the analytically computed next-to-leading
order cross section. This calculational procedure has the advantage
of reproducing the correct (leading order)
large $p_T(t\bar t )$ behavior where perturbation theory is good, and
giving the correct total cross section at next-to-leading order.
It will not correctly represent the detailed
physical low transverse momentum
behavior of the $t\bar t$ pair; however, it will give the correct
next-to-leading order integral over
the low and moderate $t\bar t$ transverse momentum. Since we are interested
in jet activity associated with $t\bar t$ production and decay, the
jet cutoffs that will be applied will mean that the
$p_T(t\bar t )$ distributions will be integrated up to the jet-cut
transverse momentum, discussed in the next section.

The parton shower method includes approximate all orders QCD corrections
to the initial and final state partons of the hard scattering subprocesses
by coupling QCD dynamics valid in the collinear approximation,
with exact kinematics, leading to explicit non-collinear multi-parton
emissions.
We denote our application of parton showers to the $2\rightarrow 2$ and
$2\rightarrow 3$
subprocesses as PS22 and PS23, respectively. As with the ME23 method,
in the PS23 Monte Carlo we use the transverse momentum cutoff
$p_{T\,c}(t\bar t )$. Since showering does not change the normalization
of the total cross section, the cross sections from PS23 and ME23 are
identical. To obtain the same normalization using the PS22 Monte Carlo, we
multiply the PS22 results by the above mentioned K-factor
(approximately equal to 1.5 for $m_t=140$~GeV) which depends on the top mass.

The initial state showering is done using
a leading-log algorithm with backwards showering. The starting (space-like)
virtuality for the initial state showers for the PS22 calculation is
taken to be $\hat{s}$,
the parton center of mass energy, unless specified otherwise.
For PS23, we take the
$\hat t$ or $\hat u$ channel virtuality defined by the difference
squared of the final state $t\bar t$ momentum with one of the
initial parton's momentum.
With this choice, the parton emission described by the hard scattering
in the $2\rightarrow 3$ matrix element is the {\it first leg} of the initial
state
shower, where the virtualities are ordered in $|t|$, and run backwards
until a cutoff virtuality $|t_c |=(3\ {\rm GeV})^2$ is reached,
whereupon the initial shower terminates.

In our final state showering algorithm, we include angular
ordering\cite{MW} effects, which model some of the soft gluon interference
effects. For PS22, we use an initial angle factor
\begin{eqnarray}
\xi_{t,\bar t} ={{p_t\cdot p_{\bar t}}\over {E_t E_{\bar t}}}
\eqnum{3}
\end{eqnarray}
and evolve down in $\xi$ until  $\xi <{{(m_t+\sqrt{|t_c |})^2}\over {E_t^2}}$,
whereupon the shower is terminated. This results in a suppression
of radiation in a cone about the direction of the
radiating heavy flavor, as given by QCD\cite{MWHF}. For the case of PS23,
we label momenta as $p_1 +p_2\rightarrow p_3 +p_4 +p_5$ where $p_5$ is the
final state light quark or gluon. Then, if $p_5$ is a gluon, we take
initial angle factors of
\begin{eqnarray}
\xi_3 &=& \xi_{35}, \cr
\xi_4 &=& \xi_{45},\ {\rm and}\cr
\xi_5 &=& min(\xi_{35} ,\xi_{45}),
\eqnum{4}
\end{eqnarray}
where $\xi_{ij}={{p_i\cdot p_j}\over {E_i E_j}}$, and $\xi_i$ is the starting
angle factor for particle $i$. If $p_5$ is a quark, we use the same
$\xi_5$, but for $\xi_3$ and $\xi_4$ adopt
\begin{eqnarray}
\xi_3 &=& \xi_4 =\xi_{34}.
\eqnum{5}
\end{eqnarray}
The final state parton showers terminate once an angle factor corresponding to
the cutoff virtuality is reached.

Top quark decay matrix elements via a real $W$ are incorporated
including $W$ spin effects, but neglecting top spin correlations, which may
be destroyed by hadronization effects. Furthermore, we terminate our
cascade decay chain at the $b$-quark level, and do not include hadronization
effects in our calculations.
The decay products of the $t$ quarks can also undergo final state
parton showers. For instance, for $t\rightarrow bW^+ \rightarrow bu\bar d$, we
choose
$\xi_{u,\bar{d}}=\xi_{u\bar d}$, whereas $\xi_b$ is determined by top quark
decay kinematics.

Major differences between the three different computational methods
can be seen in Fig. 1, where we plot the $p_T(t\bar t )$ distribution
for $m_t=140$~GeV for {\it a}) $p\bar p$ collisions at $\sqrt{s}=1.8$~TeV,
and {\it b}) $pp$ collisions at $\sqrt{s}=40$~TeV. The ME23 calculation
describes high $p_T(t\bar t )$ behavior well, since this region ought to be
dominated by one-parton emission; the low $p_T(t\bar t )$ region ought
to be dominated by multiple parton emissions, so ME23 is not expected to be
an accurate representation of reality. The PS22 result includes more reasonable
low $p_T(t\bar t )$ behavior, but does not track the ME23 result at all well
for high $p_T(t\bar t )$. (We have checked that our PS22 result agrees well
with
results generated using PYTHIA\cite{JET}.)
Finally, the merged calculation of PS23
tracks well the ME23 behavior at high $p_T(t\bar t )$, but with an overall
enhancement in the large $p_T(t\bar t )$ distribution.
The low $p_T$
region is smeared out by extra multiple parton emissions. Quantitatively, the
average $<p_T(t\bar t )>$ for Fig. 1a is given by 8, 29 and 19 GeV,
respectively, for ME23, PS22 and PS23. Thus, the lack of exact high-$p_T$
dynamics for the first parton emission in PS22 probably leads to too hard
a $p_T(t\bar t)$ spectrum. At $\sqrt{s} =40$~TeV, the PS22 again has a
different shape to the high $p_T(t\bar t)$ distribution.

\section{Results on jet activity in top pair events}

\subsection{$p\bar p$ colliders}

We first present results for the Fermilab Tevatron collider. We perform
calculations for jet activity associated with both single and dilepton events.
For single lepton plus jet events, we require
\begin{itemize}
\item $p_T(l)>20$~GeV, $|\eta (l)|<2.5$~GeV,
\item lepton isolation: in a cone of size $\Delta R=0.3$ about lepton
direction, $\Sigma |E_T|<3$~GeV,
\item $E\llap/_T >20$~GeV
\end{itemize}
For dilepton events, we relax the $p_T(l)$ cut to
\begin{itemize}
\item $p_T(l)>15$~GeV, and also require
\item $30^0<\Delta\phi (l,\bar l')< 150^0$
\end{itemize}
We also coalesce partons to form jets, using a jet cone size of
\begin{itemize}
\item $\Delta R=0.7$ and requiring
\item $p_T(jet)>15$~GeV, $|\eta (jet)|<3$.
\end{itemize}

In Fig. 2 we show the cross section in picobarns for
{\it a}) $p\bar{p}\rightarrow t\bar t\rightarrow 1l+n-$jets and
{\it b}) $p\bar{p}\rightarrow t\bar t\rightarrow 2l+n-$jets,
as a function of the number of jets $n$, at $\sqrt{s}=1.8$ TeV.
The three calculational methods have been used:
ME23, PS22 and PS23, each yielding the same total cross section.
The branching fraction of $(2\times{2\over 9}\times{6\over 9})$
has been used for $1l$
events, while $({2\over 9})^2$ has been used for dileptons.
The three methods agree to within a factor of 2
for $n(jet)\le 4$ ($n(jet)\le 2$) for
$1l$ ($2l$) events, where the jet activity is dominated by the top quark decay
products. The ME23 approach for $n(jet)=5$ (3) for $1l$ ($2l$) events lies
significantly below the shower predictions for these jet multiplicities; for
higher jet multiplicities, of course, ME23 makes no prediction.  In spite of
the differing $p_T(t\bar t)$ distributions, the PS22 and PS23 predictions
are very close for all jet multiplicities. Most of the additional jet activity
comes from initial state showers. For PS23, the initial shower virtualities
are determined by the hard-scattering subprocess, while for PS22, the initial
virtuality can be equated with any typical mass scale in leading log
approximation. Hence, we show results for PS22 for two scale choices:
$|t|=\hat{s}$ and $|t|=m_t^2$. Even for these two scale choices, the PS22
predictions differ only slightly, which may be because $\log m_t^2$ is not so
different from $\log \hat{s}$ for Tevatron energies.

To show the effect of jet multiplicity on top quark mass reconstruction,
we have examined $1l+\ge 4$ jet events, for which $W+jets$ background can
be manageable\cite{GIELE}. In Fig. 3,
we have plotted both the trijet invariant mass,
as well as associated $l\nu + jet$ mass for each of the three methods.
We have not assumed $b$-vertex identification, and that $b$'s are identified
with hadronic depositions. There is then ambiguity about which of the jets
and leptons are associated with which of the produced $t$-quarks. For this
plot,
we have added in hadronic energy smearing of $\Delta E_T =0.7\sqrt{E_T}$
on the partons. The $l$ and $\nu$ four-vectors are constrained to make up
a $W$ mass. Then 3-jet combinations and $jet-l-\nu$ combinations are
constructed; the choice for best combination gives the smallest difference
between $m(3-jet)$ and $m(jet-l-\nu )$. In Fig. 3, the ME23 approach
gives a fairly sharp peak at the top  mass value $m_t=140$~GeV. The additional
jet activity from the PS22 and PS23 approach results in a smearing caused
in part by wrong jet choices, or by parton jet broadening. The different
parton shower approaches give little difference in top mass distributions.

The main effect of parton showering is to smear out distributions and
to spread out the next-to-leading order cross-section over a large
number of $1l$ ($2l$)$+n$-jet rates.
In view of this, one should review the proposal by Berends,
Tausk and Giele\cite{GIELE} to get an indication of the top quark mass
using the relative sizes of the $2l+2$-jet and $2l+1$-jet rates, via the
jet fractions
\begin{eqnarray}
{f}_n ={\sigma (2\ell +n-jet)\over \sum_{m=0}^2 \sigma (2\ell +m-jet)}.
\eqnum{6}
\end{eqnarray}
They use the $q\bar{q}\rightarrow t\bar t \rightarrow b\bar{b}+2\ell +2\nu$
matrix element
to compute the jet rates as a function of
the top quark mass. The jet fractions for $n>2$ vanish in this
leading order approximation. The jet fractions are nearly
independent of the factorization and renormalization scales,
however they do depend on the top quark mass\cite{GIELE}.
Given that parton showering distributes the cross section over a larger
number of $2\ell+n$-jet cross sections, one might suppose that
the uncertainties due to showering could swamp any theoretical
effect due to the top quark mass.
To check, we plot in Fig. 4 the jet fractions $f_0,\ f_1$ and $f_2$ versus
the top quark mass, for the ME23 and PS23 approaches. The jet
fractions agree well with each other, and with the ME22 calculation of
Ref.~\cite{GIELE}, although slightly different cuts are used.

\subsection{$pp$ supercolliders}

We next turn to results for hadron supercolliders, with the SSC as our example.
For SSC single lepton plus jet events, we require
\begin{itemize}
\item $p_T(l)>40$~GeV, $|\eta (l)|<2.5$~GeV,
\item lepton isolation: in a cone of size $\Delta R=0.4$ about lepton
direction, $\Sigma |E_T|<0.25\times E(l)$,
\item $E\llap/_T >40$~GeV
\end{itemize}
For dilepton events, we relax the $p_T(l)$ cut to
\begin{itemize}
\item $p_T(l)>20$~GeV, and also require
\item $30^0<\Delta\phi (l,\bar l')< 150^0$
\item $E\llap/_T >20$~GeV
\end{itemize}
We coalesce partons to form jets, now using a jet cone size of
\begin{itemize}
\item $\Delta R=0.4$ and requiring
\item $p_T(jet)>30$~GeV, $|\eta (jet)|<2.5$.
\end{itemize}

In Fig. 5, we show the cross section in picobarns for
{\it a}) $pp\rightarrow t\bar t\rightarrow 1l+n-$jets and
{\it b}) $pp\rightarrow t\bar t\rightarrow 2l+n-$jets,
as a function of the number of jets $n$, at $\sqrt{s}=40$ TeV.
As in the Tevatron case, the three methods all give consistent
estimates of the low jet multiplicity cross-sections, which are dominated
by top quark decay products. In this case, however, there is much more
ambiguity
in the high jet multiplicity cross sections from PS22, for the two
different choices of initial state virtuality. This is due to the fact
that at SSC, there is a substantial cross-section for events being produced
with very large values of $\hat s$. In fact, PS22 programs should ideally be
``tuned'' in virtuality to agree with data. The PS23 approach gives larger
cross-sections than PS22 for high jet multiplicities. However, in the PS23
case,
since the initial state virtualities are specified by the hard scattering,
there
are less adjustable parameters.

There has recently been much research on heavy Higgs boson
production\cite{BARG}
and effective Lagrangian models\cite{GUN,BAG} of strong
vector boson interactions at supercollider energies.
We focus
here on heavy Higgs production, though the commentary follows as
well for effective Lagrangian models.

For a Higgs mass $m_H> 2 m_W$, the Higgs will have a large branching
fraction into $W^+W^-$ pairs. For $m_H\stackrel{>}{\sim} 500$ GeV, a large
fraction of
the Higgs production cross section and decay into W pairs comes from
$q\bar{q}\rightarrow W^+W^- q\bar{q}$. The interaction kinematics are such that
the final state partons are largely in the forward (large $|\eta|$)
region. Requiring two forward jets greatly reduces the
QCD background, mainly $q\bar{q},gg \rightarrow t\bar t \rightarrow b\bar{b}
W^+ W^-$,
however, the double tag also significantly reduces the already small
Higgs signal. A single forward jet tag does well to reduce the
QCD background from $q\bar{q}$ and $gg$ production of
$t\bar t$, however, one is left with
$gg\rightarrow t\bar t g,\ qg\rightarrow t\bar t q,\ q\bar{q}\rightarrow t\bar
t g$
processes followed by $t\rightarrow bW$.
A number of authors\cite{BARG,GUN,BAG} have found that this
background is significantly
reduced by imposing the forward jet tag and a central jet veto, which
vetos a large number of $b$-quark jets coming from top decays.
The previous analyses were done using the ME23 method.
The PS method, however,
is well-suited for evaluating multiple parton emission effects, particularly
in the forward region.

With this in mind, we show in Fig. 6 the rapidity distribution of
the highest $p_T$ jet (lower curves) as well
as the rapidity of {\it all} jets (upper curves) in $pp\rightarrow t\bar t$
events,
using the ME23, PS22 and PS23 approaches. The lower curves contain one entry
per generated event, and hence all three approaches have the same
normalization,
while the upper curves can have multiple entries per event (one per jet), and
hence have different normalizations. Examination of the lower curves for
leading jet rapidity shows that the PS curves have a broader shape,
indicating extra jet activity particularly at large $|\eta |$. However, when
one sums over {\it all} jet entries (upper curves), it is important to note
that PS approaches yield more central jet activity as well.

We have used the explicit cuts advocated in Ref.~\cite{BARG} to examine
whether $t\bar t$ background levels increase due to the additional
forward jet activity predicted by PS models. These cuts include
\begin{itemize}
\item A. $p_T(l)>100$ GeV and $|\eta (l)|<2.0$ GeV,
\item B. $3<|\eta_j (tag)|<5,\ E_j (tag)>1\ {\rm TeV},\
     p_{T_{j}}(tag)>40\ {\rm GeV}$,
\item C. $p_{T_{j}}(veto)>30\ {\rm GeV},\ |\eta_j (veto)|<3$,
\item D. $\Delta p_T(ll)>400$~GeV.
\end{itemize}
After the above cuts, we find the result of the ME23 calculation to yield
$4.3$ fb, in agreement with Ref.~\cite{BARG}, if we take
the renormalization scale equal to the
factorization scale at a value of $\sqrt{m_t^2+p_T(t)^2}$, and include jet
coalescing.

In Table 1, we show our results from ME23, PS22 and PS23 using
the above cuts, and scale choice $Q^2=\hat{s}$. The first row gives the total
$t\bar t$ cross-section, while the second row shows
results after cuts A. and D. above are applied to leptons. The ME23 yields
the largest cross-section in this case because, since the top quarks
more frequently come out back-to-back, there is a larger probability for
{\it both} leptons to have $p_T(l)>100$~GeV. In row 3, in addition, the
forward jet tag is required. In this case, the PS methods yield about
4-5 times the background that ME23 yields, which is clearly an effect of the
multiple gluon emissions from the initial state. Row 4 shows results for
the leptonic cuts, plus a central jet veto (cut C.), but no forward jet tag.
Here, the two PS calculations are significantly lower than ME23
calculation, due to the presence of additional central jet activity.
Finally, in row 5, the results of all four cuts A.-D. are shown. It is
perhaps coincidental that the combined effects of more forward jet
activity and more central jet activity in the PS calculations combine to yield
a result in close accord with the ME23 approach.

\section{Discussion and Conclusions}

In summary, we have estimated jet activity in $t\bar t$ events by a
merging of matrix element methods and parton showers. Such a calculation
considers the light quark or gluon emission of the $2\rightarrow 3$ subprocess
to be an exact representation of the first leg of a parton shower. Our approach
gives the correct shape to the high $p_T(t\bar t)$ distribution while
including the effects of multiple parton emission, which are important
especially in the forward region. This method is complementary to fixed-order
tree level calculations\cite{ELSEX,STANGE} in that it includes
approximate all-orders QCD corrections, instead of finite-order exact
results.

We have applied our method to $p\bar p\rightarrow t\bar tX$ production relevant
to
the Tevatron top search experiments. The PS23 approach verifies ME23
jet multiplicity results for low jet multiplicity, and also verifies
jet multiplicity estimates using the more common PS22 approach. Top
quark mass distributions suffer more smearing using the parton shower
approaches, which is to be expected due to greater uncertainty in matching
jets with partons, and jet broadening effects.

For supercollider energies, the PS23 approach yields higher jet
multiplicities than the PS22 or ME23 approach. Since the PS
calculation is most valid for multi-parton emissions in the forward
region, it ought to be used for $t\bar t$ background estimates
that depend on forward jet tagging. In our calculations, we find
that PS methods lead not only to more forward jet activity, but also to more
central jet activity, than ME23 calculations. The ME23 calculation has
been used by a number of authors\cite{BARG,GUN,BAG}
to estimate backgrounds to $WW$ events
from heavy Higgs decays or strong $WW$ scattering.

We have advocated here the PS23 method, incorporating the $2\rightarrow 3$ hard
scattering matrix element merged with parton showering. This method correctly
reproduces the large $p_T(t\bar t)$ behavior at leading order and
provides a well defined scale at which to begin the initial state showers.
The ultimate test of the method, however, will come from a comparison
with experimentally measured distributions.

%
\acknowledgements
We thank K. Cheung, S. Protopopescu and A. Stange for discussions.
This research was supported in part by the U.~S. Department of Energy under
contract number DE-FG05-87ER40319, and the National Science Foundation,
under grant number PHY-9104773.
\newpage
%
%
%
%

\newpage
%
%
\figure{Transverse momentum distribution of $t\bar t$ pair for three
calculations described in the text for
{\it a}) $p\bar p$ collisions at $\sqrt{s}=1.8$~TeV and {\it b})
$pp$ collisions at $\sqrt{s}=40$~TeV, for $m_t =140$ GeV.
\label{FIG1}}
%
\figure{Cross-section as a function of jet multiplicity for $p\bar p$
collisions at $\sqrt{s}=1.8$~TeV for {\it a}) $1l+n$-jet events, and
{\it b}) $2l+n$-jet events using cuts described in the text,
for $m_t =140$ GeV.
\label{FIG2}}
%
\figure{Invariant mass plots of jet-lepton-neutrino system, and three jet
system for $p\bar p$ collisions at $\sqrt{s}=1.8$~TeV, using three
calculational methods, for $m_t =140$ GeV.
\label{FIG3}}
%
\figure{Jet fractions versus top quark mass for $p\bar p$
collisions at $\sqrt{s}=1.8$~TeV.
\label{FIG4}}
%
\figure{Cross-section as a function of jet multiplicity for $pp$
collisions at $\sqrt{s}=40$~TeV for {\it a}) $1l+n$-jet events, and
{\it b}) $2l+n$-jet events using cuts described in the text,
for $m_t =140$ GeV.
\label{FIG5}}
%
\figure{Pseudo-rapidity distribution for highest $p_T$ jets (lower curves)
and each jet in $pp\rightarrow t\bar tX$ events, at $\sqrt{s}=40$~TeV, for {\it
a})
$1l+n$-jet events, and
{\it b}) $2l+n$-jet events using cuts described in the text, except
$|\eta_j |<5$, for $m_t =140$ GeV.
\label{FIG6}}
\newpage
%
\begin{table}
\caption{Cross-section from $pp\rightarrow t\bar tX$ in fb after cuts given in
text,
at $\sqrt{s}=40$~TeV, for $m_t =140$ GeV.}
\begin{tabular}{cccc}
cuts & ME23 & PS22 & PS23 \\
\tableline
no cuts & $1.1\times 10^7$ & $1.1\times 10^7$ & $1.1\times 10^7$ \\
A,D     & 1567 & 727 & 1266 \\
A,B,D   & 100. & 393 & 567  \\
A,C,D   & 13.2 & 3.0 & 6.2  \\
A,B,C,D & 2.5  & 2.0 & 2.0  \\
\end{tabular}
\label{TABLE1}
\end{table}
%
%
%
\end{narrowtext}
%

\begin{references}
\bibitem{MTRAD} The LEP Collaborations, Phys. Lett. {\bf 276B}, 247 (1992).
\bibitem{BBP} For a review of top quark signatures, see H. Baer, V. Barger
and R. Phillips, Phys. Rev. {\bf D39}, 3310 (1989).
\bibitem{CDF} F.~Abe {\it et. al.} , (CDF Collaboration), Phys. Rev. Lett.
{\bf 66}, 447 (1992); see also talk by N. Shaw, Workshop on Physics at Current
Accelerators and the Supercollider, Argonne, IL, (1993).
\bibitem{ELSEX} R. K. Ellis and J. Sexton, Nucl. Phys. {\bf B282}, 642 (1987).
\bibitem{NDE} P.~Nason, S. Dawson and R. K. Ellis, Nucl. Phys. {\bf B303}, 607
(1988).
\bibitem{BEEN} W. Beenakker, H. Kuijf, W. van Neerven and J. Smith,
Phys. Rev. {\bf D40}, 54 (1989); W. Beenakker, W. van Neerven, R. Meng,
G. Schuler and J. Smith, Nucl. Phys. {\bf B351}, 507 (1991).
\bibitem{NDET} P.~Nason, S. Dawson and R. K. Ellis,
Nucl. Phys. {\bf B327}, 49 (1988).
\bibitem{NASON} M. Mangano, P. Nason and G. Ridolfi, Nucl. Phys. {\bf B373},
295
(1992).
\bibitem{ISA} F. Paige and S. Protopopescu, in {\it Supercollider Physics},
Proceedings of the Workshop on Super High Energy Physics, Eugene, Oregon,
1985, ed. D. E. Soper (World Scientific, Singapore, 1986), p. 41.
\bibitem{JET} T. Sjostrand, Computer Phys. Comm., {\bf 39}, 347 (1986);
T. Sjostrand and M. Bengtsson, {\it ibid.}, {\bf 43}, 367 (1987);
H. U. Bengtsson and T. Sjostrand, {\it ibid.}, {\bf 46}, 43 (1987).
\bibitem{HER} B. Webber and G. Marchesini, Nucl. Phys. {\bf B310}, 461 (1988);
G. Marchesini {\it et. al.}, Cavendish-HEP-90/26 (1991).
\bibitem{BR} H. Baer and M. H. Reno, Phys. Rev. {\bf D44}, R3375 (1991) and
Phys. Rev. {\bf D45}, 1503 (1992).
\bibitem{BARG} V. Barger, K. Cheung, T. Han and D. Zeppenfeld,
Phys. Rev. {\bf D44}, 2701 (1991) and Fermilab-Pub-93/092-T.
\bibitem{GUN} D. Dicus, J. Gunion, L. Orr and R. Vega,
Nucl. Phys. {\bf B377}, 31 (1991).
\bibitem{BAG} J. Bagger {\it et. al.}, Fermilab-Pub-93/040-T (1993).
\bibitem{HMRS} P. N. Harriman, A. D. Martin, W. J. Stirling and R. G. Roberts,
Phys. Rev. {\bf D42}, 798 (1990).
\bibitem{MW} G. Marchesini and B. Webber, Nucl. Phys. {\bf B238}, 1 (1984); for
a review, see B. Webber, Ann. Rev. Nucl. Part. Sci. 1986 {\bf 36}, 253 (1986).
\bibitem{MWHF} G. Marchesini and B. Webber, Nucl. Phys. {\bf B330}, 261 (1990).
\bibitem{GIELE} F.A. Berends, J.B. Tausk and W.T. Giele,
Phys. Rev. {\bf D47}, 2746 (1993).
%
\bibitem{STANGE} see A. Stange (unpublished) for tree level
${\cal O}(\alpha_s^2 )$ corrections to $t\bar t$ production.
%
\end{references}
\end{document}